\documentclass[twocolumn,pra,aps,showpacs]{revtex4}
\usepackage{xspace,amsmath,amsfonts,amsthm,amssymb,amsbsy,graphicx,color,setspace}
\pdfoutput=1

\begin{document}

\def\ra{\rangle}
\def\la{\langle}
\def\vp{\varphi}
\def\be{\begin{equation}}
\def\ee{\end{equation}}
\def\ba{\begin{eqnarray}}
\def\ea{\end{eqnarray}}
\def\ap{\alpha}
\def\bt{\beta}
\newcommand{\bra}[1]{\left\langle #1 \right\vert}
\newcommand{\ket}[1]{\left\vert #1 \right\rangle}
\newcommand{\re}[1]{\langle #1 \rangle}
\newcommand{\bx}{\begin{matrix}}
\newcommand{\ex}{\end{matrix}}
\newcommand{\mc}[1]{\mathcal}

\title{Parity Measurements in  Quantum Optical Metrology}
\author{Aravind Chiruvelli and Hwang Lee}

\affiliation{Hearne Institute for Theoretical Physics, Department of Physics and Astronomy, Louisiana State University, Baton Rouge, LA 70803, USA}

%\date{\today }
\begin{abstract}
We investigate the utility of parity detection to achieve Heisenberg-limited phase estimation for optical interferometry.  We consider the parity detection with several input states that have been shown to exhibit sub shot-noise interferometry with their respective detection schemes. We show that with parity detection, all these states achieve the sub-shot noise limited phase estimate. Thus making the parity detection a unified detection strategy for quantum optical metrology.  We also consider quantum states that are a combination of a NOON states and a dual-Fock state, which gives a great deal of freedom in the preparation of the input state, and is found to surpass the shot-noise limit. 
\end{abstract}
\maketitle

\section{Introduction}
Quantum optical metrology deals with the estimation of an unknown phase by exploiting the quantum nature of the  input state under consideration~\cite{kok04,seth:me}. Due to the inherent uncertainty imposed by quantum mechanics, the problem reduces to minimizing the uncertainty of the expectation value of a suitable observable~\cite{holl,Bollinger:PR,bou,gerr,dunin,leib,poo,resc,uys,durk,brien,dorner,yurke,rosetta,yuen,lane,optimal,hradil,noh,berry,pezze,naga,sun,huver}.  In the usual classical setting, for a given mean number of input photons, $\bar{N}$, the phase estimate scales as $1/\sqrt{\bar{N}}$, which is usually referred as shot-noise limit (SL). It has been shown that by exploiting the signature quantum properties such as entanglement the uncertainty can be reduced to the Heisenberg limit (HL) of $1/{\bar{N}}$; an enhancement of a factor of $\sqrt{\bar{N}}$. Achieving a sub-shot noise limit or the Heisenberg limit depends on the nature of the input states and the detection strategy of the output measurement~\cite{brien,rosetta,naga}.%~\cite{lane,optimal,hradil,noh,berry,pezze,naga,sun,huver}. 

Precise optical phase measurement has been an open problem for many years and has many applications, most notably the gravitational wave detection \cite{caves,kimb}. 
Various input states of light and measurement schemes have been shown to surpass the SL. Sanders and Milburn have computed the optimal measurement~\cite{optimal}, written as Positive Operator Valued Measure (POVM), to achieve the HL. The method specified by Sanders and Milburn is independent of the system phase and thus an optimal one. Berry and Wiseman  have considered the optimal POVM and derived the optimal input state to achieve the minimum uncertainty that scales as the HL~\cite{berry}. They also showed a way of approximately implementing the optimal POVM using a feedback technique. With this feedback technique along with the Kitaev algorithm for phase estimation \cite{QCtext}, Higgins {\it et al.} have experimentally achieved the HL scaling of the optical phase measurement~\cite{nature}. 

{In this article, we will discuss the parity measurement, which detects whether the number of photons in a given output mode is even or odd. The scope of the article is restricted to the local phase estimation as opposed to the global~\cite{durk,hradil}, such that the phase estimation can be evaluated by using the linear error propagation formula\cite{text}. The primary objective of this paper is to investigate the utility of parity detection scheme and not the role of the quantum correlations--also referred as entaglement--in achieving the sub--shot--noise limited phase estimate. The precise question on the role of entanglement in achieving sub--shot--noise limited interferometry is beyond the scope of this article.The curious reader may find Ref.~\cite{maur} useful in that direction.}

As we make extensive use of the Schwinger representation to analyze the Mach Zehender Interferometer (MZI), we wish to present a brief discussion of the representation. Any four-port optical lossless device, such as the MZI considered here, can be conveniently described using the Schwinger representation of the angular momentum~\cite{yurke}. The operators, which form an SU(2) rotation group, and describe the MZI \cite{text} are: $\hat{J}_x=(\hat{a}^\dagger\hat{b}+\hat{a}\hat{b}^\dagger)/2, \hat{J}_y=(\hat{a}^\dagger\hat{b}-\hat{a}\hat{b}^\dagger)/2i, \hat{J}_z=(\hat{a}^\dagger\hat{a}-\hat{b}^\dagger\hat{b})/2$, where $\hat{a}$ and $\hat{b}$ are the mode operators which obey bosonic commutation relation, $[\hat{a},\hat{a}^\dagger]= [\hat{b},\hat{b}^\dagger]=1$. The angular momentum operators obey $[\hat{J}_i,\hat{J}_j]=i\epsilon_{ijk}\hat{J}_k$. The total photon number is $\hat{N}=\hat{a}^\dagger\hat{a}+\hat{b}^\dagger\hat{b}$, and $\hat{J}^2=\hat{J}_{x}^2+\hat{J}_{y}^2+\hat{J}_{z}^2=(\hat{N}/2)(\hat{N}/2+1)$ is the Casimir invariant. The generator for beam-splitter transformation is usually represented by $\hat{J}_x$~\cite{optimal,noh}. The combined two mode input state is represented by the simultaneous eigen state of $\hat{J}^2$ and $\hat{J}_z$, i.e $|j\mu\rangle_z$, where $|j+\mu\rangle$ and $|j-\mu\rangle$ represent $|N\rangle_a$ and $|N\rangle_b$ respectively and $j=N/2$ for a fixed input photon number $N$. Correspondingly, if $|j\nu\rangle_z$ represents the output state, then $\nu$ represents the output photon number difference, $(N_a-N_b)|_{out}$. In this representation the Mach Zehender Interferometer is given by an operator, $\hat{\mathcal{I}}={\rm exp}({-i\varphi\hat{J}_y})$. For a given input state $\ket{j\mu}$, the output state can be written as, $\hat{\mathcal{I}}\ket{j\mu}=  e^{-i\varphi\hat{J}_y}\ket{j\mu}=\sum_{\nu=-j}^j d_{\nu,\mu}^j(\vp)\ket{j\nu}$, where $ d_{\nu,\mu}^j(\vp)$ is the usual rotation matrix elements~\cite{rot,ang}.

The paper is organized as follows. In section \ref{PD} we discuss the parity measurement and setup a general framework of calculating the expectation value for an arbitrary input state. In section \ref{app} we apply it to specific input states including a combination of a NOON state and a dual-Fock state, followed by section \ref{conc} with conclusions.

\begin{figure}[htp]
\includegraphics[height=0.75\hsize,width=1.0\hsize]{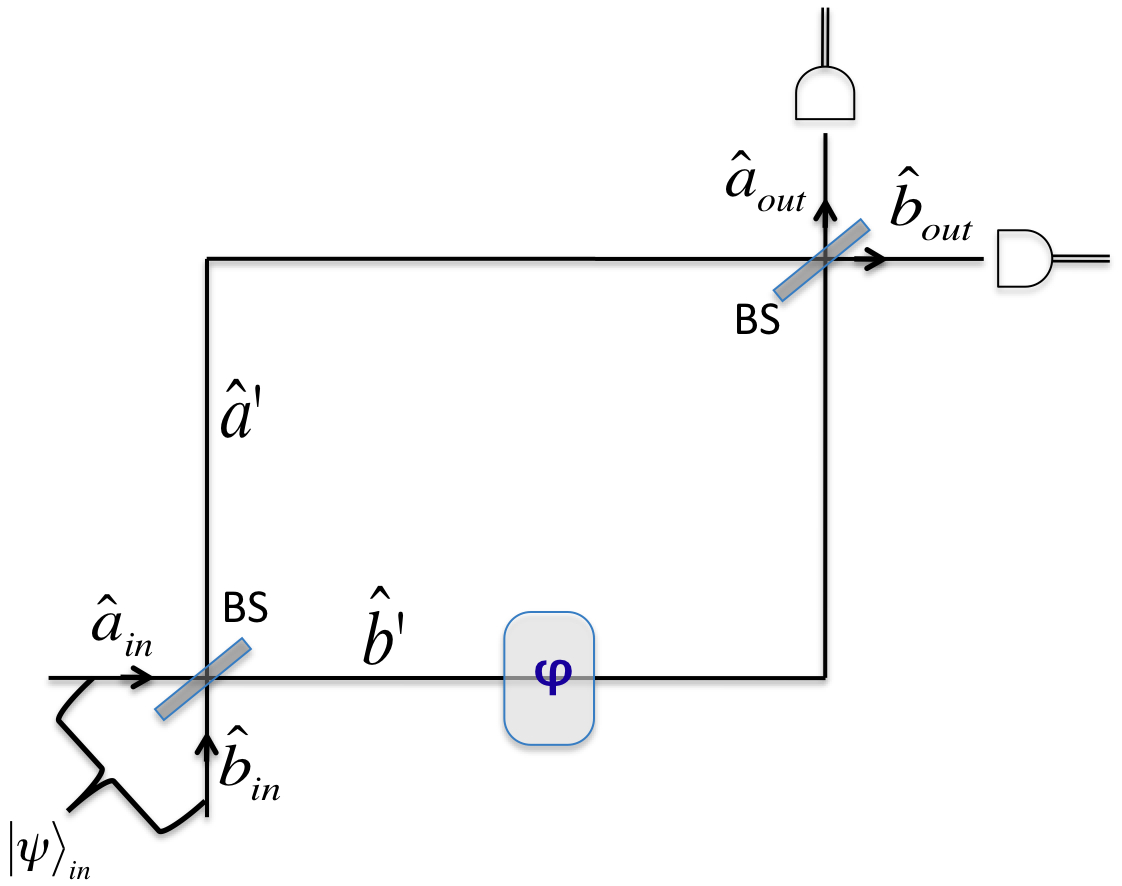}
\caption{Schematic of a Mach-Zehender Interferometer (MZI). $|\psi\rangle_{in}$ represents the joint input state at $\hat{a}_{in}$ and $\hat{b}_{in}$.The photon number states entering at $\hat{a}_{in},(\hat{b}_{in})$ are $|N\rangle_a(|N\rangle_b)$. The symbol $\varphi$ represents the relative phase between the modes, within the interferometer. BS: Beam Splitter.}
\label{fig1}
\end{figure}

\section{Parity Detection}
\label{PD}

 Parity detection was first proposed by Bollinger {\it et al.} in 1996 to study spectroscopy with a  ``maximally entangled'' state of trapped ions~\cite{Bollinger:PR}. The detection considered there is $(-1)^{\hat{N}_{e}}$, where $\ket{N}_{e}$ represents $N$ atoms in excited state. It is straightforward to draw  parallel between the two level atom and the MZI depicted in Fig.~\ref{fig1}. In the case of the optical phase measurements, we have $\ket{\psi}_N$ the NOON state. 
Formally, in the present context of phase measurement, considering the MZI shown in Fig.~\ref{fig1}, the NOON states are written as~\cite{sand, boto,kok,fiur,walt,mitch,guo,pryde03}:
\begin{equation}
|\psi\rangle_N = \frac{|N\rangle_{a'}|0\rangle_{b'}+|0\rangle_{a'}|N\rangle_{b'}}{\sqrt{2}},
\label{noon}
\end{equation}
where $N$ is the number of photons and $a'$ and $b'$ are the two internal modes of the MZI.

The detection operator which was first proposed by Gerry~\cite{gerr}:
 \be
 \hat{P}= (-1)^{\hat{b}_{out}^\dagger\hat{b}_{out}}=(-1)^{j-\hat{J}_z}.
 \label{op}
 \ee
 
 There is no particular reason to perform such a detection on $\hat{b}_{out}$ and choosing $\hat{a}_{out}$ works equally well. Gerry and Campos have applied this operation to interferometry with the NOON states, resulting in the exact HL~\cite{cnoon}. As stated in the introduction, these states are not the input states of the MZI, but are states {\it after} the first beam--splitter (BS). One can quite easily write down the input state using a beam--splitter transformation as we will do it now. In the Schwinger representation, the NOON states is represented as: $\ket{\psi}_N=(\ket{j,j}+\ket{j,-j})/\sqrt{2}$. Then we get the actual input state:
 \be
 \ket{\psi}_i=e^{-i\frac{\pi}{2}\hat{J}_x}\ket{\psi}_N =\sum_{\mu=-j}^jA_{\mu}\ket{j,\mu},
 \label{inoon}
 \ee
 where the coefficient $A_{\mu}$ is given by
 \be
 A_{\mu}=\frac{1}{\sqrt{2}}\left[e^{i(\mu-j)\pi/2}d_{\mu,j}^j(\frac{\pi}{2})+e^{i(\mu+j)\pi/2}d_{\mu,-j}^j(\frac{\pi}{2})\right].\nonumber
 \ee
 
\begin{figure}[htp]
\includegraphics[height=0.5\hsize,width=0.8\hsize]{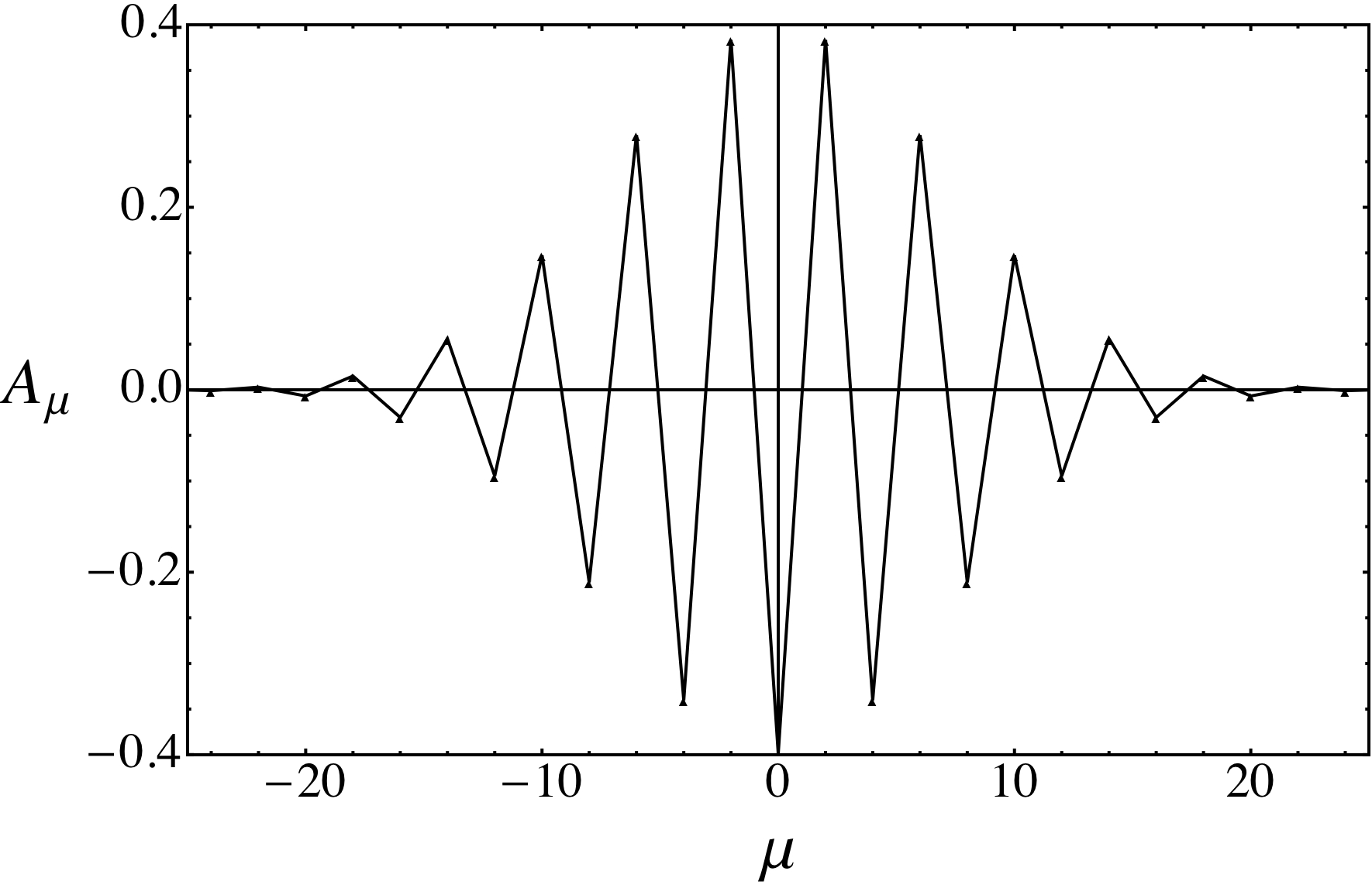}
\caption{The coefficients of the state given in Eq.~(\ref{inoon}) for an input photon number, $N$=100. This is the input state to the MZI shown in Fig.~\ref{fig1} such that it is NOON state {\it after the first beam splitter} and thereby giving the Heisenberg-limited phase estimate.}
\label{noonco}
\end{figure}

We used $e^{-i\frac{\pi}{2}\hat{J}_x}=e^{i\frac{\pi}{2}\hat{J}_z}e^{-i\frac{\pi}{2}\hat{J}_y}e^{-i\frac{\pi}{2}\hat{J}_z}$ in obtaining the above result. Thus $\ket{\psi}_i$ is the input state of the MZI to get the exact HL with parity detection. The real part of the coefficients $A_{\mu}$ are plotted in Fig.~\ref{noonco}.

Recently, Uys and Meystre~\cite{uys} noted that a state whose coefficients look alike the coefficients plotted in Fig.~\ref{noonco} also gives an exact Heisenberg limit. They obtained this result via numerical simulations and an explicit mathematical expression of the input state was not given. From Fig.~\ref{noonco} it appears that the state obtained in Ref.~\cite{uys} is a beam--splitter transformation of the NOON state which is given by Eq.~(\ref{inoon}). 

For a local phase estimation the phase uncertainty is typically given as~\cite{text}:
 \be 
 \delta\varphi=\frac{\Delta\hat{P}}{|\partial_\vp\langle\hat{P}\rangle|},
 \label{vari}
 \ee 
  where $\Delta\hat{P} = \sqrt{\la\hat{P}^2\ra-\la\hat{P}\ra^2}=  \sqrt{1-\la\hat{P}\ra^2}$, since $\hat{P}^2=1$. 
  
The usual approach in phase estimation one wishes to take, is to fix the detection strategy, i.e. to fix a particular phase dependent observable at the output and look for the behavior of the input states. While a detection of the signal is done at the output ports, one can always transform the observable through the second beam splitter and can think of the detection {\it within} the interferometer. Obviously, both descriptions are equivalent but the latter may help understanding the direct effect of the measurement on the states after acquiring the phase shift. 

We can obtain more insight on $\hat{P}$ by transforming through the BS and further express it as a projection operator by using the completeness relation:
 \ba
 \hat{Q}=e^{-i\frac{\pi}{2}\hat{J}_x}\hat{P}e^{i\frac{\pi}{2}\hat{J}_x} = (-1)^je^{i\pi\hat{J}_y} \;\;\;\;\;\;\ \nonumber & & \\=\sum_{\nu,\mu=-j}^j(-1)^jd^{j}_{\nu,\mu}(-\pi)|j\nu\rangle\langle j\mu|.\;\;\;\;\;\ \label{pq} \ea It is straight forward to see that $\hat{Q}^2=1$. In terms of the photon numbers, and noting that $d^{j}_{\nu,\mu}(-\pi)=(-1)^{2\nu}\delta_{\nu,-\mu}$, we get~\cite{rot}
 \begin{equation}
\hat{Q}=i^N\sum_{k=0}^N(-1)^{k} |k,N-k\rangle\langle N-k,k|. 
\label{in}
\end{equation} 

If the NOON states are under consideration for the above expression of the observable $\hat{Q}$, the only relevant terms would be for $k=0$ and $k=N$. Thus the observable considered in Refs. \cite{rosetta, huver,mark,gilb} is: $\hat{Q}_N=\ket{0,N}\bra{N,0}+\ket{N,0}\bra{0,N}$, that leads to the HL for NOON states. This gives the same expectation value, and thus $\hat{Q}_N$ can be implemented with parity detection for the NOON states. Explicitly, after the phase shifter, $\ket{\psi}_N$ becomes: $\ket{\psi}_{N,\varphi}=\ket{N,0}+e^{iN\varphi}\ket{0,N}$, which leads to (see also Ref.~\cite{yang}):
\be
\bra{\psi}\hat{Q}\ket{\psi}_{N,\varphi}=\bigg \{ \bx i^{N+1} \sin N \varphi , & \;\;\ N \;\ \text{odd}, \\
i^N \cos N \varphi , & \;\;\ N \;\ \text{even}. \ex 
\label{Qo}
\ee
Using Eq.~(\ref{Qo}) and Eq.~(\ref{vari}) we immediately get  $\delta\vp=1/N$.

Now we shall obtain the expectation value of parity observable for an arbitrary input state. In the Schwinger notation an arbitrary two-mode input and the corresponding output states are written as,
\ba
 \ket{\psi}_{\text{in}}= \sum_{2j=0}^\infty \sum_{\mu=-j}^j \psi_{\mu,j}\ket{j,\mu}, \;\;\;\;\;\;\;\;\;\;\;\;\;\;\;\;\;\;\;\;\;\;\;\;\ \\
 \ket{\psi}_{\text{out}}=\hat{\mathcal{I}}\ket{\psi}_{\text{in}}=\sum_{2j=0}^\infty \sum_{\mu=-j}^j \psi_{\mu,j}e^{-i\varphi\hat{J}_y}\ket{j,\mu}, 
 \ea
where $\psi_{\mu,j}$ is the amplitude for the arbitrary input state. Using the above equation it is straightforward to calculate the expectation value of $\hat{P}$ for an arbitrary input state
\ba
\la\hat{P}\ra_{\rm out} =~_{\rm out}\la{\psi}|\hat{P}\ket{\psi}_{\rm out} \;\;\;\;\;\;\;\;\;\;\;\;\;\;\;\;\;\;\;\;\;\;\;\;\;\;\;\;\;\;\;\;\;\;\;\;\;\;\;\;\;\;\;\;\;\;\;\ \nonumber \\  = \sum_{2j=0}^\infty\sum_{\mu,\mu'=-j}^j(-1)^{j-\mu'}\psi_{\mu',j}^{*}\psi_{\mu,j}d_{\mu'\mu}^j(2\vp).\;\;\;\;\;\;\;\;\;\;\;\
\label{res}
\ea

In obtaining the above result, we used  $\bra{j'\mu'}e^{i\vp\hat{J}_y}e^{-i\pi\hat{J}_z}e^{-i\vp\hat{J}_y}\ket{j\mu}=\sum_{\nu=-j}^j(-1)^{-\nu}d_{\nu,\mu'}^j(\vp)d_{\nu,\mu}^j(\vp)\delta_{j,j'}$. 
The summation over $2j$ describes the situation where the number of photons are not fixed. In what follows we will use a fixed number of input photons $N$, and this summation will then be dropped.  
  
\begin{figure*}[tp]
\includegraphics[width=1.0\hsize,height=0.25\hsize]{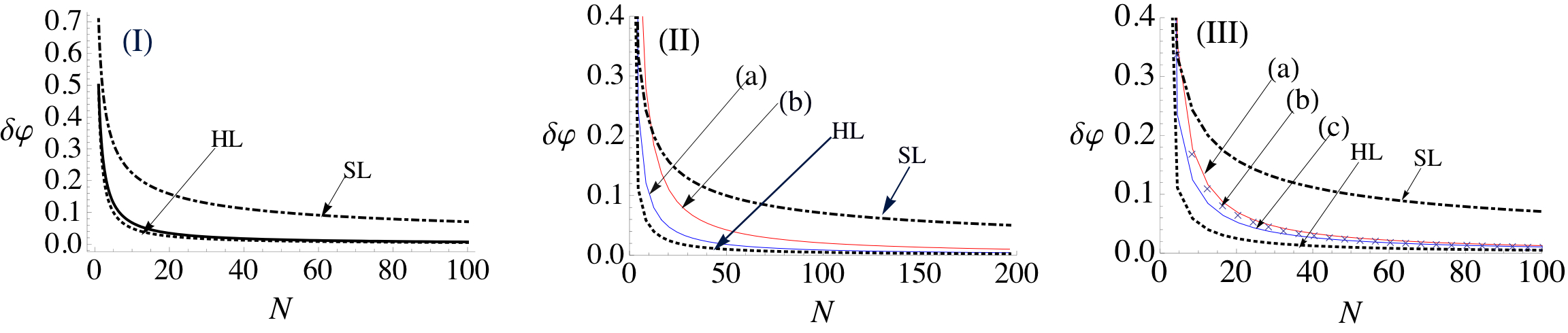}
\caption{The variance of $\delta \varphi$ versus $N$, the input number of photons. (I) The dual fock state, shown as the continuous curve which is same as given in Ref.~\cite{cdual}. (II) For the combined state $\ket{\psi}_{\rm c}$, in the absence of relative phase, where (a) $\alpha=\sqrt{\frac{2}{3}}, \beta =\sqrt{\frac{1}{3}}, \theta=0$, (b) $\alpha=\sqrt{\frac{1}{3}}, \beta =\sqrt{\frac{2}{3}}$ and $\theta=0$. (III) For a fixed $\alpha=\beta=\frac{1}{\sqrt{2}}$, (a) $\theta =0$, (b) represented by $\times$ for $\theta =\pi$, (c) $\theta =\frac{\pi}{4}$. In (III) the plots for the case (a) and (b), almost overlap. In all plots the dot-dashed curve is the shot-noise limit (SL) and the dotted curve represents the Heisenberg limit (HL). }
\label{dual}
\end{figure*}

\section{Application of the parity detection}
\label{app} 
\subsection{Parity detection with uncorrelated states}
We begin with a coherent state input at mode $\hat{a}_{in}$. We have at the input:
\be
\ket{\psi}_\alpha=\sum_{2j=0}^\infty e^{-\frac{|\alpha|^2}{2}}\frac{(\alpha)^{2j}}{\sqrt{(2j)!}}\ket{j,j},
\ee
where $|\alpha|=\bar{n}$, the average photon number. Using Eq. (\ref{res}), we have:
\ba
\la\hat{P}\ra_\ap= \sum_{2j=0}^\infty e^{-\frac{|\alpha|^2}{2}}\frac{(|\alpha|^2)^{2j}}{(2j)!}d_{j,j}^j(2\vp) \;\;\;\;\;\;\;\;\;\;\;\ \nonumber \\
=\text{exp}\left[-|\ap|^2+\frac{|\ap|^2\sqrt{1+\cos(2\vp)}}{\sqrt{2}}\right],
\ea
which, in the limit  $\vp\rightarrow0$, according to Eq. (\ref{vari}), immediately leads to $\delta\vp_\ap=1/{\sqrt{\bar{n}}}$. We thus recover the shot-noise limit. This can also be obtained by $\hat{J}_z$ measurement at the output, which corresponds to the photon number difference. However, in the case of  a $\hat{J}_z$ measurement, the shot-noise limit is reached when $\vp$ tends to odd multiples of $\pi/2$~\cite{text}. 

The next simplest uncorrelated state is a number state at $\hat{a}_{in}$ and vacuum at $\hat{b}_{in}$. Thus the input state: $\ket{\psi}_s=\ket{N}_a\ket{0}_b=\sum_{\mu=-j}^j\delta_{\mu,j}\ket{j,j}$. The subscript $s$ denotes the input at a single port of MZI. Using Eq. (\ref{res}), we obtain:
\be
\la\hat{P}\ra_s=\frac{\left[1+\cos(2\vp)\right]^j}{2^j},
\ee
for which, in the limit $\vp\rightarrow0$, we get $\delta\vp_s=1/\sqrt{2j}=1/\sqrt{N}$. This result shows that parity detection gives the same result as the $\hat{J}_z$ measurement for a single-port input state~\cite{jon98}. 

Now we consider the dual-Fock input state~\cite{holl,poo,noh,sun}. Campos {\it et al.} have shown the utility of the parity measurements for the dual Fock input state~\cite{cdual}. Their analysis also includes a comparison and contrast of the quantum state distribution with the interferometer with the NOON states (as these are the states within the interferometer). The dual Fock state can be written as: $\ket{\psi}_d=\ket{N,N}=\sum_{\mu=-j}^j\delta_{\mu,0}\ket{j\mu}$. Using Eq.~(\ref{res}) we immediately get $\la\hat{P}\ra_d=(-1)^j d_{0,0}^j(2\vp)$. Using  $\la\hat{P}\ra_d$ in Eq.~(\ref{vari}) for a small phase, $\vp\rightarrow0$, we get $\delta\vp_d={1}/\sqrt{2j(j+1)}\propto {1}/{N}$ (see also Ref.~\cite{yang}). We plot this in Fig.~\ref{dual}(I), which is the same as shown in Ref.~\cite{cdual}.

\subsection{Parity detection with a combined state}

As discussed above, both the dual-Fock state and the NOON state, have an Heisenberg limited phase variance with the parity detection. 

It is natural to ask how precisely a state has to be prepared to take the advantage of parity or, in general, any detection scheme. We now attempt to answer this question by considering a combination of a NOON state [see Eq.~(\ref{inoon})] and a dual-Fock state such as 
\ba
\ket{\psi}_{\rm c}={\rm C}_N(\alpha\ket{\psi}_i+\beta\ket{\psi}_d)\nonumber \;\;\;\;\;\;\;\;\;\;\;\;\;\;\;\;\;\;\;\;\;\;\;\;\;\ \\
={\rm C}_N\left(\alpha\frac{e^{-i\frac{\pi}{2}\hat{J}_x}\left(\ket{j,j}+\ket{j,-j}\right)}{\sqrt{2}}+\beta\ket{j,0}\right).
\ea
Writing $\ap=|\ap|e^{i\theta_\ap}$ and $\bt=|\bt|e^{i\theta_\bt}$, where the normalization constant is given by 
\be
{\rm C}_N=[1+2\sqrt{2}|\alpha||\beta|d_{j,0}^j(\pi/2)\cos(\theta-\frac{N\pi}{4})]^{-\frac{1}{2}},\ee where $\theta=\theta_{\ap}-\theta_{\bt}$ is the relative phase. With such a quantum state as the input of the MZI in Fig.~\ref{fig1}, we have the output $\ket{\psi}_{\rm c|out}\equiv\mathcal{I}\ket{\psi}_{\rm c}$. It gives rise to:
\ba
\bra{\psi}\hat{P}\ket{\psi}_{\rm c|out}={\rm C}_N^2 \lbrace|\ap|^2\la\hat{P}\ra_{\rm NOON} +|\bt|^2\la\hat{P}\ra_{\rm dual} \nonumber \\ +i^j2\sqrt{2}|\ap||\bt|d_{0,0}^j\cos(N\vp)\cos(\theta) \rbrace, \;\;\;\
\\
\text{ where} \;\;\;\;\;\;\;\;\;\;\;\;\;\;\;\;\;\;\;\;\;\;\;\;\;\;\;\;\;\;\;\;\;\;\;\;\;\;\;\;\;\;\;\;\;\;\;\;\;\;\;\;\;\;\;\;\;\;\;\;\;\;\;\;\;\;\;\;\ \nonumber \\
\la\hat{P}\ra_{\rm NOON}=(-1)^j(e^{iN\vp}+(-1)^Ne^{-iN\vp})/2, \nonumber \;\ \\
\la\hat{P}\ra_{\rm dual}= (-1)^jd_{0,0}^j(2\vp). \nonumber \;\;\;\;\;\;\;\;\;\;\;\;\;\;\;\;\;\;\;\;\;\;\;\;\;\;\;\
\label{sup}
\ea
The above result is obtained using the symmetric properties of the rotation matrix elements~\cite{rot} and the Baker-Campbell-Hausdorff (BCH) formula~\cite{text}. We can now use this in Eq.~(\ref{vari}) to calculate $\delta\vp$. To gain more insight, we will first take $\theta=0$ and look for the $\delta\vp$ for various combinations of $|\ap|$ and $|\bt|$ using $|\ap|^2+|\bt|^2=1$. The results are shown in the top portion of Fig. \ref{dual}(II). 

Next, we fix $|\ap|=|\bt|=1/\sqrt{2}$ and vary $\theta$. This is plotted in the in the bottom portion of Fig. \ref{dual}(III). In both the cases we can clearly see that with $\ket{\psi}_{\rm c}$ at the input, using parity detection, we do get the sub-shot noise limited variance of the optical phase. This result implies a wider applicability of the parity detection. 
%FIGURE
\begin{figure*}[ht]
\includegraphics[width=1.0\hsize,height=0.25\hsize]{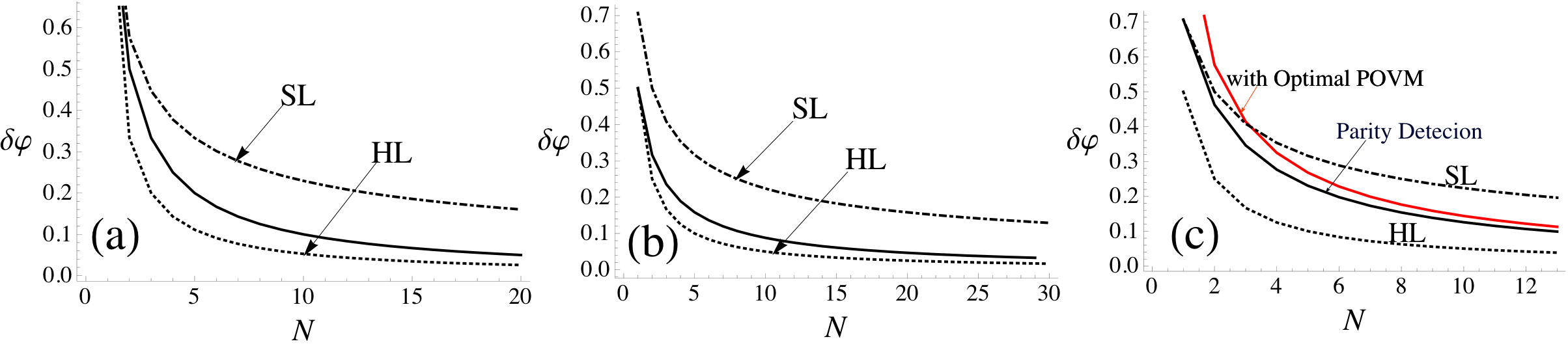}
\caption{The variance $\delta \vp$  as a function of $N$, the input number of photons for various states. In all the figures, the dotted line corresponds to the Heisenberg limit(HL) and dot-dashed to the shot-noise limit(SL). The continuous line in (a) for the ``modified"-Yuen state, (b) for the $\ket{\psi}_{\rm sp}$ considered in Ref.~\cite{pezze} and (c) for the optimal state(for comparison $\delta\vp $ as obtained with\cite{berry} optimal POVM is also shown) }
\label{fig4}
\end{figure*}
%END FIGURE

\subsection{Parity detection with correlated states} 
{  In this section we consider the quantum-correlated states that have been already shown to achieve sub-shot noise limited phase estimate under several different detection schemes. Here we show that all these states does achieve the goal with the parity detection and thus making the parity detection, a unified detection scheme}. 

Let us begin with the Yurke state. The Yurke state is formally written as $[|N/2\rangle_a|N/2\rangle_b+|(N+1)/2\rangle_a|(N-1)/2\rangle_b]/\sqrt{2}$. Yurke {\it et al.}~\cite{yurke} have shown that by using the output photon difference, a minimum phase sensitivity of $2/N$ can be achieved.  In the Schwinger notation Yurke state is  $\ket{\psi}_Y=\sum_{\mu=-j}^j\frac{1}{\sqrt{2}}(\delta_{\mu,0}+\delta_{\mu,1})\ket{j,\mu}$. Using Eq.(\ref{res}) we get the expectation value of parity as~\cite{yang}:
\be
\la\hat{P}\ra_Y=\frac{(-1)^j}{2}[d_{0,0}^j(2\vp)-d_{1,1}^j(2\vp)+2d_{0,1}^j(2\vp)].
\ee
Again in the limit $\vp\rightarrow0$, we have using Eq. (\ref{vari}) we get: $\delta\vp\rightarrow1/\sqrt{j(j+1)}\propto\sqrt{2}/N$, which is same as that obtained with the $\hat{J}_z$ measurement~\cite{yurke}.

Let us now consider another correlated input state first proposed by Yuen~\cite{yuen}: 
\be
\ket{\psi}_{\rm yu}=\sum_{\mu=-j}^j\frac{1}{\sqrt{2}}(\delta_{\mu,1/2}+i\delta_{\mu,-1/2})\ket{j,\mu}.
\ee
Now by using Eq.~(\ref{vari}), we get: $\la\hat{P}\ra_{\rm yu}=0$. Thus the parity detection does not give any phase information for the Yuen state. The main reason for the  vanishing $\la\hat{P}\ra_{\rm yu}$ is the relative phase of $\pi/2$ among the two possible inputs at the MZI. This motivates us to consider the state with zero relative phase, which would be a slightly modified form of $\ket{\psi}_{\rm yu}$. Let us define the ``modified" Yuen state as,
 \be
 \ket{\psi}_{\rm m.yu}=\sum_{\mu=-j}^j\frac{(\delta_{\mu,1/2}+\delta_{\mu,-1/2})}{\sqrt{2}}\ket{j,\mu},
 \ee
 which then following Eq.~(\ref{res}) leads to:
\be
\re{\hat{P}}_{\rm m.yu}=i(-1)^j d_{\frac{1}{2},-\frac{1}{2}}^j(2\vp).
\ee
Thus using Eq.~(\ref{vari}) one can calculate the variance. In the limit of $\vp\rightarrow 0$ this leads to a sub-shot noise variance as is shown in Fig. \ref{fig4}(a).
%BEGIN TABLE
\begin{table*}[htp]
\caption{ Phase estimate for various input states under parity detection. Here, Row 1: Coherent state at one input ($\hat{a}_{in}$). Row 2: Number state at one input ($\hat{a}_{in}$). Row 3: Dual Fock state. Row 4: Yurke state. Row 5: ``Modified" Yuen state. Row 6: State suggested by Pezze and Smerzi. Row 7: Berry-Wiseman Optimal state. Row 8: NOON state. The states in 7 and 8 are states representing the modes $\hat{a}'$ and $\hat{b}'$ {\it within} the interferometer. In the Schwinger notation, $\hat{J}_y$ eigen states represent the internal modes $\hat{a}'$ and $\hat{b}'$(see Ref.~\cite{noh}). }
\begin{tabular}{ c | c | c || c }
  \hline
  \multicolumn{3}{ c }{Input State} & {Phase Estimate}\\
  \hline
   & Fock- state Notation & Schwinger Notation & $\delta\phi$ \\
  \hline
 \hline
  1. & $\ket{\alpha}_a\ket{0}_b$ & $\displaystyle\sum_{2j=0}^\infty e^{-\frac{|\alpha|^2}{2}}\frac{(\alpha)^{2j}}{\sqrt{(2j)!}}\ket{j,j}$  & $\delta\vp= \frac{1}{\sqrt{\bar{N}}} ({\rm SL})$ \\ \hline

  2. & $\ket{N}_a\ket{0}_b$& $\displaystyle\sum_{\mu=-j}^j\delta_{\mu,j}\ket{j,j}$ &$ \delta\vp\rightarrow\frac{1}{\sqrt{N}}$ \\  \hline

  3 & $\ket{N}_a\ket{N}_b$  & $\displaystyle\sum_{\mu=-j}^j\delta_{\mu,0}\ket{j\mu}$ & $\delta\vp =\frac{\sqrt{2}}{\sqrt{N(N+2)}}\approx\frac{\sqrt{2}}{N}$\\ 
  \hline

  4. & $\frac{1}{\sqrt{2}}\left[{\ket{\frac{N}{2}}_a\ket{\frac{N}{2}}_b+\ket{\frac{N}{2}+1}_a\ket{\frac{N}{2}-1}_b}\right]$ & $\displaystyle\sum_{\mu=-j}^j\frac{(\delta_{\mu,0}+\delta_{\mu,1})}{\sqrt{2}}\ket{j,\mu}$ & $\delta\vp\rightarrow\frac{1}{\sqrt{\frac{N}{2}(\frac{N}{2}+1)}}\approx\frac{\sqrt{2}}{N}$\\ \hline

   5. & $\frac{1}{\sqrt{2}}\left[{\ket{\frac{N+1}{2}}_a\ket{\frac{N-1}{2}}_b+\ket{\frac{N-1}{2}}_a\ket{\frac{N+1}{2}}_b}\right]$ & $\displaystyle\sum_{\mu=-j}^j\frac{(\delta_{\mu,1/2}+\delta_{\mu,-1/2})}{\sqrt{2}}\ket{j,\mu}$ & sub-shot noise [Fig.~\ref{fig4}(a)]\\ \hline

   6. & $\frac{1}{\sqrt{2}}\left[{\ket{\frac{N}{2}+1}_a\ket{\frac{N}{2}-1}_b+\ket{\frac{N}{2}-1}_a\ket{\frac{N}{2}+1}_b}\right]$ & $\displaystyle\sum_{\mu=-j}^j\frac{(\delta_{\mu,1}+\delta_{\mu,-1})}{\sqrt{2}}\ket{j,\mu}$ & sub-shot noise [Fig.~\ref{fig4}(b)] \\ \hline
   
   7. & $\sqrt{\frac{2}{N+1}}\sum_{k=0}^N \sin\left[\frac{(N+2+k)\pi}{2(N+2)}\right]\ket{\frac{N+k}{2}}_{a'}\ket{\frac{N+k}{2}}_{b'}$& $ \displaystyle\sum_{\mu=-j}^j\frac{\sin(\frac{(\mu+j+1)\pi}{2j+2})}{\sqrt{j+1}}\ket{j,\mu}_y$ & sub-shot noise [Fig.~\ref{fig4}(c)] \\
    \hline
 
   8. & $\frac{1}{\sqrt{2}}{\ket{N}_{a'}\ket{0}_{b'}+\ket{0}_{a'}\ket{N}_{b'}}$ & $\displaystyle\sum_{\mu=-j}^j\frac{(\delta_{\mu,j}+\delta_{\mu,-j})}{\sqrt{2}}\ket{j,\mu}_y$ & $\delta\vp= \frac{1}{N}$ ({\rm HL}) \\

\hline
\end{tabular}
\label{table}
\end{table*}

%END TABLE

Berry and Wiseman have the so-called proposed optimal states~\cite{berry}, for the use of the optimal POVM proposed by Sanders and Milburn~\cite{optimal}. We now consider this state in the context of the present work. Formally, the optimal state is~\cite{berry}: 
\be
\ket{\psi}_{\rm opt}=\sum_{\mu=-j}^j {\rm C}_{\mu}\ket{j,\mu},
\ee where the amplitude is given by
 \be
 {\rm C}_{\mu}=\frac{1}{\sqrt{j+1}}\sin\left[\frac{(\mu+j+1)\pi}{2j+2}\right]. \nonumber
 \ee 

It is worth noting that the optimal state is a state {\it within} the interferometer just like the case of the NOON state. So we use the operator $\hat{Q}$ as the detection operator. It would be a simple BS transformation to obtain the actual MZI input state and this result is given in Ref.~\cite{berry}. First, we need to transform this state through the phase shifter: $\ket{\psi}_{\vp|{\rm opt}}=\sum_{\mu=-j}^je^{-i\mu\vp}{\rm C}_{\mu}\ket{j,\mu}$. Then, using Eq.~(\ref{pq}) we obtain,
\be
\la\hat{Q}\ra_{\rm \vp|opt}=\sum_{\nu=-j}^j(-1)^{2\nu}{\rm C}_{-\nu}{\rm C}_{\nu}e^{i2\nu\vp}.
\label{opt}
\ee

Plugging the above equation into Eq.~(\ref{vari}) we can calculate $\delta\vp$. However, we  have not found  a closed form expression for $\delta\vp$. Instead, Fig. \ref{fig4}(b) shows a numerical plot. Clearly the optimal state gives a sub-shot noise level and does better for large photon number. 
It is worth noting that, although one can generate the optimal state, the implementation of the optimal POVM~\cite{berry} for which $\ket{\psi}_{\rm opt}$ is designed for, requires a real time feedback. The variance in this case is~\cite{berry}: 
\be
\delta\vp_{opt}=\tan\left(\frac{\pi}{N+2}\right)\approx\frac{\pi}{N}.
\ee 
The parity detection, on the  contrary, is relatively straightforward. The variance thus obtained due to optimal POVM and parity detection for the $\ket{\psi}_{\rm opt}$ is shown in Fig. \ref{fig4}(b).\\

Finally we consider the input state, recently suggested by Pezze and Smerzi~\cite{pezze} for achieving the HL, given by 
\be
\ket{\psi}_{\rm ps}=\frac{(\ket{j,1}+\ket{j,-1})}{\sqrt{2}}.
\label{sp}
\ee

The strategy employed by these authors is direct detection of number of photons at the output modes of MZI and applying Bayesian analysis for multiple detections with greater confidence. Also it is important to note that Eq.~(\ref{vari}) was not used to calculate the variance but a single interferometric measurement was used. In this sense, it is claimed~\cite{pezze} that the above state is the most optimal one for the  HL. Such a quantum state of Eq. (\ref{sp}) was also considered in Ref~\cite{jon98} by Dowling with the $\hat{J_z}$ measurement. 

Here we apply parity detection to the input state $\ket{\psi}_{\rm ps}$. Using Eq. (\ref{res}), we immediately get the following result:
\be
\re{\hat{P}}_{\rm ps}=(-1)^{j+1} (d_{1,1}^j(2\vp)+d_{-1,1}^j(2\vp)),
\ee
and the phase variance can be calculated using Eq. (\ref{vari}). We plot the result numerically Fig.~\ref{fig4}(b), in the limit of $\vp\rightarrow 0$. And we clearly see the sub-shot noise limit. Indeed for the large number of input photons, the phase estimate approaches the HL.

\section{Conclusions}
\label{conc}
To summarize, in this article, we demonstrate the importance of the parity detection scheme in the optical phase estimation. By considering a combination of the NOON state and the dual-Fock state at the input, we have shown that the parity detection still gives sub-shot noise variance, and it reaches close to the HL for large input number of photons. We have also considered the Yuen state, Pezze and Smerzi optimal state, the Berry-Wiseman optimal state and have shown that we can achieve even smaller phase variance using the parity detection. Our results indicate that the parity detection acts as a unified detection scheme for precision phase measurements. We have summarized the results in the Table~\ref{table}.
\begin{figure}[htp]
\includegraphics[height=0.4\hsize,width=0.65\hsize]{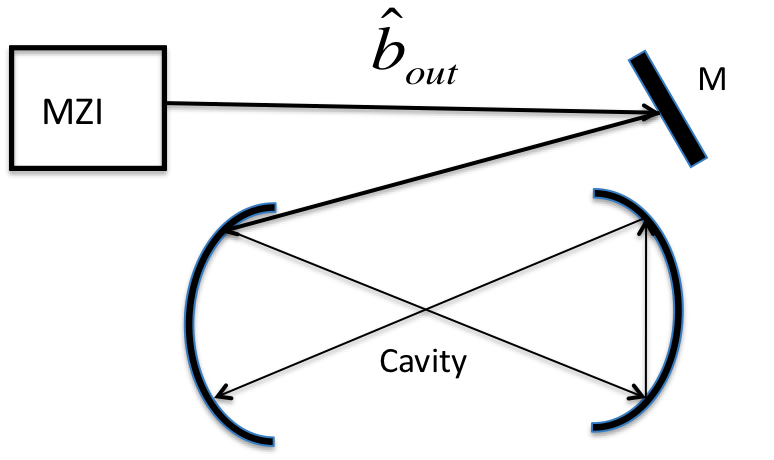}
\caption{Schematic of a containing the detecting output mode $\hat{b}_{out}$ from the MZI in Fig.~\ref{fig1} in a cavity towards measuring the parity of the mode as suggested in Ref.~\cite{lutter}. The output mode is reflected at mirror, M, and then fed into the cavity. By properly picking the curvature of the cavity morrors, one can trap the mode for sufficiently long time.}
\label{fig5}
\end{figure}

{ From a mathematical point of view, the parity detection appears to be simple in comparison with any other detection strategy, but to experimentally realize such a detection scheme is not trivial. There are basically two different approaches to accomplish the task in laboratory. The first one is by employing number-resolving photodetectors~\cite{cdual,dary,hwlee,ware,rosen,eqo,lita08,afek09} at the output detecting mode (mode $\hat{b}_{out}$ in Fig.~\ref{fig1}). It should be noted that we do not necessarily need a photodetector at single photon resolution, rather a detector that would discriminate even and odd number of photons would suffice~\cite{gerry05}.

 Alternatively, one can measure the Wigner function. In Ref.~\cite{royer} it has been shown that expectation value of the parity is $h/2$ times the  Wigner function. In Ref.~\cite{lutter}, it has been shown that if we can store the single mode field in a cavity, then we can perform the parity detection by measuring the Wigner function. This would require containing the output detecting mode $\hat{b}_{out}$ in a cavity as shown in Fig.~\ref{fig5}. By properly picking the curvature of the cavity morrors, one can trap the mode for sufficiently long time. This approach is discussed further in Ref.~\cite{spring}.}

\section*{\bf Acknowledgments}
This work is supported by the Intelligence Advanced Research Projects Activity, the Army Research Office, and the Defense Advanced Research Projects Agency. We would like to acknowledge J. P. Dowling and Y. Gao for stimulating discussions.


\begin{thebibliography}{13}
%1
\bibitem{kok04}
P. Kok, S. L. Braunstein, and J. P. Dowling
J. Opt. B: Quantum Semiclass. Opt. 6 S811-S815 (2004)

%2
\bibitem{seth:me}
V. Giovannetti, S. Lloyd, and L. Maccone
Phys. Rev. Lett. {\bf 96}, 010401 (2006)
%2
\bibitem{holl}
M. J. Holland and K. Burnett, Phys. Rev. Lett. {\bf 71}, 1355 (1993).
%3
\bibitem{Bollinger:PR}
J. J.\ Bollinger, W. M.\ Itano, D. J.\ Wineland, and D. J.\ Heinzen,
Phys. Rev. A {\bf{54}}, R4649 (1996).
%4
\bibitem{bou}
P. Bouyer and M. A. Kasevich, Phys. Rev. A {\bf 56}, R1083 (1997).
%5
\bibitem{gerr}
C. C. Gerry, Phys. Rev. A {\bf 61},043811 (2000).
%6
\bibitem{dunin}
J. A. Dunningham, K. Burnett, and S. M. Barnett, Phys. Rev. Lett. {\bf 89}, 150401 (2002).
%7
\bibitem{leib}
D. Leibfried {\it et al.}, Science {\bf 304}, 1476 (2004).
%8
\bibitem{poo}
R. C. Pooser and O. Pfister, Phys. Rev. A {\bf 69}, 043616 (2004).
%9
\bibitem{resc}
K. J. Resch {\it et al.}, Phys. Rev. Lett., {\bf 98}, 223601 (2007).
%10
\bibitem{uys}
H. Uys and P. Meystre, Phys. Rev. A {\bf 76}, 013804 (2007).
%11
\bibitem{durk}
G. A. Durkin and J. P. Dowling, Phys. Rev. Lett., {\bf 99}, 070801 (2007).
%12
\bibitem{brien}
J. L. OÕBrien, Science, {\bf 318}, 1393 (2007).
%13
\bibitem{dorner}
U. Dorner {\it et al.}, 
Phys. Rev. Lett., {\bf 102}, 040403 (2009).
%14
\bibitem{yurke}
B.\ Yurke, S. L.\ McCall, and J. R.\ Klauder,
Phys. Rev. A {\bf {33}}, 4033 (1986).

%15
\bibitem{rosetta}
H.\ Lee, P.\ Kok, and J.P.\ Dowling, 
J. Mod. Opt. {\bf {49}}, 2325 (2002).
%16
\bibitem{yuen}
H. P.\ Yuen,
Phys. Rev. Lett {\bf {56}}, 2176 (1986).
%17
\bibitem{lane}
A. S. Lane, S. L. Braunstein, and C. M. Caves, Phys. Rev. A {\bf 47}, 1667 (1993).
%18  
\bibitem{optimal}
B. C. Sanders and G. J. Milburn, Phys.Rev. Lett. {\bf 75}, 2944 (1995).
%19
\bibitem{hradil}
Z. Hradil, Phys. Rev. A {\bf 51}, 1870 (1995).
%20 
\bibitem{noh}
T. Kim, {\it et al.}, Phys. Rev. A {\bf 57}, 4004 (1998).
%21 
\bibitem{berry}
D. W. Berry and H. M. Wiseman, Phys. Rev. Lett. {\bf 85}, 5098 (2000).
%22 
\bibitem{pezze}
L. Pezze and A. Smerzi, Phys. Rev. A {\bf 73}, 011801(R) (2006).
%23
\bibitem{naga}
T. Nagata {\it et al.}, Science, {\bf 316}, 726 (2007).
%24
\bibitem{sun}
F. W. Sun {\it et al.}, Europhys. Lett. {\bf 82}, 24001 (2008).
%25
\bibitem{huver}
S. D. Huver, C. F. Wildfeuer, and J. P. Dowling, Phys. Rev. A, {\bf 78}, 063828 (2008).

%26
\bibitem{caves}
C. M.\ Caves, Phys. Rev. D {\bf 23}, 1693 (1981).

%27
\bibitem{kimb}
H. J. Kimble {\it et al.}, Phys. Rev. D {\bf 65}, 022002 (2001).
%28
\bibitem{sand}
B. C. Sanders, Phys. Rev. A, {\bf 40}, 2417 (1989).
%29
\bibitem{boto}
A. N. Boto {\it et al.}, Phys. Rev. Lett., {\bf 85}, 2733 (2000).
%30 
\bibitem{cnoon}
C. C. Gerry and R. A. Campos, Phys. Rev. A {\bf 64},063814 (2001).
%31
\bibitem{kok}
P. Kok, H. Lee, and J. P. Dowling, Phys. Rev. A, {\bf 65}, 052104 (2002).
%32
\bibitem{fiur}
J. Fiurasek, Phys. Rev. A, {\bf 65}, 053818 (2002).

%33
\bibitem{walt}
P. Walther {\it et al.}, Nature, {\bf 429}, 158 (2004).
%34
\bibitem{mitch}
M. W. Mitchell, J. S. Lundeen, and A. M. Steinberg, Nature, {\bf 429}, 161 (2004).
%35
\bibitem{guo}
F. W. Sun, Z.Y. Ou, and G. C. Guo, Phys. Rev. A {\bf 73}, 023808 (2006).
%new
\bibitem{pryde03}
G. J. Pryde and A. J. White
Phys. Rev. A,  {\bf 68}. 052315 (2003)
%36
\bibitem{leibf}
D. Leibfried et al., Nature, {\bf 438}, 639 (2005).
%37 
\bibitem{mark}
M. A. Rubin and S. Kaushik, Phys. Rev. A, {\bf 75}, 053805 (2007).
%38 
\bibitem{gilb}
G. Gilbert, M. Hamrick, and Y. S. Weinstein, J. Opt. Soc. Am. B {\bf 25}, 1336 (2008).
%39
\bibitem{text}
C. C.\ Gerry and P. L.\ Knight, {\it Introductory Quantum Optics}, 
(Cambridge University Press, Cambridge, UK, 2005).
%40
\bibitem{QCtext}
M. A. Nielsen and I. L. Chaung, {\it Quantum computation and Quantum Information}, (Cambridge University Press, Cambridge, UK, 2000).
%41
\bibitem{nature}
B. L.\ Higgins {\it et al.},
Nature {\bf{450}}, 393 (2007).

%42
\bibitem{rot}
The rotation matrix element: \newline$\langle j\nu|e^{-i\theta\hat{J}_y}|j\mu\rangle=d^{j}_{\nu\mu}(\theta)=(-1)^{\nu-\mu}2^{-\nu}\sqrt{\frac{(j-\nu)!(j+\nu)!}{j-\mu)!(j+\mu)!}}\times P^{(\nu-\mu,\nu+\mu)}_{j-\nu}(cos\theta)(1-cos\theta)^{\frac{\nu-\mu}{2}}(1+cos\theta)^{\frac{\nu+\mu}{2}}$ where $P_n^{(\alpha,\beta)}(x)$ is the Jacobi polynomial. These rotation matrix elements further possess symmetric relations: \newline  $d^{j*}_{\nu,\mu}=d^{j}_{\nu,\mu}=(-1)^{\nu-\mu}d^{j}_{\mu,\nu}=d^{j}_{-\mu,-\nu};$ and a contraction rule: $\sum_{\nu=-j}^{j}d^{j}_{\alpha\nu}(\theta_1)d^{j}_{\nu\mu}(\theta_2)=d^{j}_{\alpha\mu}(\theta_1+\theta_2)$.

%43
\bibitem{ang}
D. M.\ Brink and G. R.\ Satchler, {\it Angular Momentum},
(Clarendon Press, Oxford, UK, 1993).

%44
\bibitem{yang}
Y. Gao and H. Lee
J. Mod. Opt. {\bf 55}, 3319 (2008).
%45
\bibitem{jon98}
J. P.\ Dowling, 
Phys. Rev. A {\bf{57}}, 4736 (1998).
%46
\bibitem{cdual}
R. A.\ Campos, C. C.\ Gerry, and A.\ Benmoussa,
Phys. Rev. A {\bf {68}}, 023810 (2003).
%47
\bibitem{eqo}
H. A. Bachor and T. C. Ralph, {\it A Guide to Experiments in Quantum Optics}, (Wiley-VCH 2004).
%48
\bibitem{dary}
D. Achilles {\it et al.}, J. Mod. Opt. {\bf 51}, 9-10, 1499 (2004).
%49
\bibitem{hwlee}
H. Lee {\it et al.}, J. Mod. Opt. {\bf 51}, 9-10, 1517 (2004).
%50
\bibitem{ware}
M. Ware {\it et al.}, J. Mod. Opt. {\bf 51}, 9-10, 1549 (2004).
%51
\bibitem{rosen}
D. Rosenberg {\it et al.},
Phys. Rev. A {\bf 71}, 061803(R), (2005).
%52
\bibitem{lita08}
A. E. Lita, A. J. Miller, S. W. Nam,
Opt. Exp. {\bf 16}, 3032 (2008).

\bibitem{afek09}
I. Afek, A. Natan, O. Ambar, and Y. Silberg,
Phys. Rev. A, {\bf 79}, 043830 (2009).


\bibitem{gerry05}
C. C. Gerry, A. Benmoussa, and R. A. Campos,
Phys. Rev. A, {\bf 72}, 053818 (2005).
%53
\bibitem{maur}
G. Mauro D'Ariano {\it et al.}. Phys. Rev. Lett. {\bf 87}, 270404 (2001).

%54
\bibitem{royer}
A. Royer,
Phys. Rev. A {\bf 15}, 449 (1977).
%55
\bibitem{lutter}
L. G. Lutterbach and L. Davidovich,
Phys. Rev. Lett., {\bf 78}, 2547-2550 (1997). 
%56
\bibitem{spring}
Chapter 6 in {\it Quantum State Estimation}, 
Springer Verlag, Series: Lecture Notes in Physics , Vol. 649,  
Edited by: Paris, Matteo; Rehacek, Jaroslav.




\end{thebibliography}
\end{document}